\documentclass[12pt]{article}

\begin{document}

\title{Structure and Reactions with Exotic Nuclei \large{within the  {\bf INFN-PI32} network.}}

\author {Angela Bonaccorso  \footnote{In collaboration with   G. Blanchon and   A. Garcia- Camacho,  INFN Sez. di  Pisa;   M. Colonna,  M. Di  Toro,  Cao  Li  Gang,  U. Lombardo,  J. R.izzo,   INFN- LNS;    S. Lenzi,  P. Lotti,  A. Vitturi,  INFN Sez. di   Padova.}\\
{\small INFN, Sez. di Pisa,} \\
{\small  and  Dipartimento di Fisica, Universit\`a di Pisa} \\ {\small  Largo Pontecorvo 3,}\\  {\small  56127 Pisa, Italy.}\\ 
{\small E-mail: bonac@df.unipi.it}}

\maketitle

\begin{abstract}
The INFN (Italian National Institute for Nuclear Physics) has approved a national theoretical network on "Structure and Reactions with Exotic Nuclei".
The project involves the INFN branches of Laboratorio Nazionale del Sud, Padova and Pisa.
The aim of the project is to start coordinating and  to homogenize the research already performed in Italy in this field and to strengthen and improve the Italian contribution on the international scenario. Furthermore it aims at  creating a solid theoretical structure to support future experimental facilities at the INFN national laboratories such as SPES at LNL and EXCYT at LNS. A review of present and future activities is presented.
\end {abstract}

\section{Introduction}

Since a few years an increasing number of Italian theoreticians has concentrated his research on the study of exotic nuclei. Such activities have so far been carried out within pre-existing national projects related to a wide spectrum of themes of nuclear dynamics, structure and reactions using many body techniques, shell model, collective modes and semiclassical or fully quantum mechanical approaches to peripheral and central reactions such as transfer and breakup, fusion, elastic scattering via microscopic optical potentials, multifragmentation.

The goal of our project is to start coordinating and homogenizing such efforts to improve our mutual understanding, and to strengthen the Italian contribution on the international scenario. Furthermore our efforts will help creating a solid theoretical structure to support future experimental activities at the INFN national laboratories.

In fact, in the last two decades, the use of radioactive beams of rare isotopes in several laboratories around the world (REX-ISOLDE at CERN, GANIL in France, GSI in Germany, CRC, Louvain la Neuve in Belgium, RIKEN in Japan, DUBNA in Russia, Argonne, MSU, Oak Ridge, Notre Dame in USA , etc.) has provided new research directions and an increasing number of researchers all over the world is converging on such subject. The INFN in Italy is also heavily involved in this field. The facility EXCYT and the large acceptance spectrometer called MAGNEX are being completed at Laboratorio Nazionale del Sud. On the other hand the first step of the SPES project at the Laboratorio Nazionale di Legnaro has been approved in the form of a proton driver. Furthermore the INFN is promoting the new European Radioactive Beam Facility (EURISOL).
Members of our collaboration are actively participating in NuPECC working groups, in particular in the preparation of "The Physics Case" for EURISOL, whose report is available at
http://www.ganil.fr/eurisol/Final\_Report/A-Physics-Case-20-Dec-02.pdf, and in general of the NuPECC Long Range Plan.

The relatively new subject of exotic nuclei is of fundamental importance because while all existing theories for the nuclear interaction and the many body nuclear structure have been based on the study of stable nuclei, very little is known about the way in which standard nuclear models work for the description of unstable nuclei with anomalous N/Z ratio. Important questions to answer are for example: the isospin dependence of the effective nuclear interaction, the modification of the traditional shell sequence with possible vanishing of the shell gaps, the persistence of collective features, the properties of nuclear matter at very low density, the form of the EOS for asymmetric nuclear matter. Similarly in the field of nuclear reactions still open questions are the identifications of the most important reaction channels and the clarification of the associated reaction mechanisms. Many of these features are also related to nuclear reactions of astrophysical interest such as those governing the primordial nucleosyntesis.

The proposed research activity will deal with the following aspects: reaction mechanisms and structure information extraction for nuclei close to the driplines, single particle and collective degrees of freedom, dynamical symmetries at the phase transitions, dynamics of heavy nuclei with anomalus N/Z ratios and isospin degrees of freedom, equation of state.
The partecipants have complementary competences in the fields of structure and reaction theory. They have common national (LNS,LNL) and international collaborations (ie : IPN, Orsay; GANIL, Caen, France, MSU, USA, etc.). Their present abilities and activities in the above research fields are described in the following.

\section{Reaction Mechanisms by the Pisa Group}

In Pisa there is a longstanding tradition for studying
peripheral reactions such as transfer and breakup,
therefore it has been easy and natural to tourn our
attention to the study of halo nuclei [1-8].

In recent years we have concentrated on a consistent
treatment of nuclear and Coulomb breakup and
recoil effects treated to all
orders and including interference effects. We have developed
a formalism
which allows the calculations of energy, momentum and
angular distributions for the core and halo particle and
absolute cross sections. The possibility of calculating
so many observables is almost unique to our model. The
dependence on the final
state interaction used has been clarified and also the
accuracy of the eikonal model compared to fully
quantum mechanical theories has been established.

An extension of the method to proton breakup has been
recently presented and we plan to apply it to the
study of reaction of astrophysical interest such as those involving $^{8}$B.
Finally a microscopic model for
the calculation of the optical potential in the breakup
channel has been developed. The method originally
used to calculate elastic scattering of halo projectiles
on ligth targets is now being extended to heavy targets
by the inclusion of recoil effects. Also we are extending
our techniques to the calculations of angular correlations.

In the last period we have started to study nuclei unbound
against neutron emission, such as $^{10}$Li and $^{13}$Be.
They are the constituents of two neutron halo nuclei
(i.e. $^{11}$Li and $^{14}$Be).
The study of their low lying resonance states is of
fundamental importance for the understanding of
two neutron halo nuclei.
The final goal is to clarify the structure of the core-neutron
interaction. This is by no means a trivial task as such cores
are themselves unstable nuclei ( $^{9}$Li, $^{12}$Be) and therefore
cannot be used as target in experimental studies.
We are at present discussing the differences between the
technique of projectile fragmentation and of transfer to the
continuum in order to understand whether they would
convey the same structure information. This line
of research is leading us naturally to study
the structure of few-body systems which we
plan to undertake in a near future.

\section {Reaction Mechanisms and Structure of Rare Isotopes by the Padova Group}

The Padova group has similar and complementary lines of
research as the Pisa group as far as reaction mechanisms are
concerned. However it has a special interest
for a somehow lower energy domain where fusion and the
coupling to breakup channels are particularly important [9-24].
Besides it is active in studying structure problems such as:

- Study of the pairing correlations in low-density
nuclear systems, as in the external part of halo nuclei.

- Microscopic estimate of inelastic excitation to
the low-lying continuum dipole strength via
microscopic continuum RPA calculations.

- Study of isospin symmetry in low- and high-spin
states in medium-mass N=Z nuclei up to $^{100}$Sn. Study of the
interplay of T=0 and T=1 pairing.

- Study of nuclear structure with algebraic models. This line of research
is associated with the use of algebraic models, as the Interacting Boson
Model or its variations, to describe different aspects of nuclear spectra.
Our traditional approach is based on the use of the concept of boson
intrinsic state. In this framework we will study the new symmetries E(5)
and X(5) associated with phase transitions
and individuate mass region far from stability where such critical
points may occur.

- Study of the role of continuum-countinuum
coupling in the break-up of weakly-bound nuclei.

\section {Isospin Dynamics in Reactions with Exotic Beams at LNS}

Two teams are active at the LNS. One is working in
the energy range from the Coulomb barrier (Tandem) to the Fermi energies
(Superconducting Cyclotron). Our main motivation is to extract
physics information on the isovector channel of the nuclear interaction
in the medium from dissipative collisions in this energy range using the
already available stable exotic ions and in perspective the new
radioactive facilties. We have developed very reliable microscopic transport
models, in a extended mean field frame, for the simulations of the reaction
dynamics in order to check the connection between the tested effective
interactions and the experiments, in particular for the isospin degree of
freedom [25-39]. This work is of interest for the understanding of the physics
behind the reaction mechanisms and for the selection of observables
most sensitive to different features of the nuclear interaction.
Moreoever we have a more general theoretical activity on the
isospin dynamics in nuclear liquid-gas phase transitions. New instabilities
have been evidenced with a different "concentration" between the gas and
cluster phases, leading to the Isospin Distillation effects recently
observed in experiments. A quantitative analysis can give direct
information on the density dependence of the symmetry term for
dilute asymmetric matter, i.e. around and below saturation. We remind
the poor knowledge of the isovector part of the nuclear effective interaction
at low densities, which is actually of large interest even for structure
calculations of drip line nuclei.

The main results obtained in the last year are related to:

1) Isospin dynamics in low energy dissipative collisions.

2) Isospin in Nuclear fragmentation.

\section{ Finite Nuclear Systems in Brueckner Theory at LNS}

The second team at LNS is interested in relating nuclear properties to
elementary interactions between nucleons and to
build up an energy density functional starting from a more
fundamental level than the present phenomenological energy
functionals of non-relativistic mean field or RMF [40-46]. This can be
achieved because of the familiarity of the group with the
Brueckner theory in infinite nuclear matter including 2-body and
3-body forces. It has been shown that the inclusion of 3-body
forces in the Brueckner theory is necessary for obtaining the
correct saturation point of nuclear matter and going away from the
so-called Coester line. From the results of infinite matter one
will construct an energy density functional which can give the
same results in nuclear matter and also can be used in finite
nuclei. This is quite similar to the
energy functional method of atomic physics based on ab
initio calculations of the homogeneous electron gas and the local
density approximation (LDA). This nuclear energy functional should
be trustable away from the stability region since no adjustment
will be made to reproduce the properties of stable nuclei,
contrarily to phenomenological energy functionals whose extrapolations can
be questionable.

The proposed method is a simpler alternative than direct Brueckner
calculations of finite systems. It also allows for studies of
excitations of nuclei, within RPA-type of
calculations built on top of the mean field ground state. This is again
in the same spirit as the time-dependent LDA (TDLDA) method
which has proved very successful in atomic cluster physics.
The main objectives of the project are:

- BHF calculations of asymmetric and polarized matter.

- Construction of the energy functional.

- Ground states of finite nuclei.

- Excitations of finite nuclei.

- Neutron star crust

\section{Conclusions}

We have presented the main research lines of the new INFN-PI32 theory network on exotic nuclei. They span from low energy reaction theory for elastic scattering and fusion, to intermediate energies studies for breakup and multi-fragmentation for the understanding of the  isospin degree of freedom. Structure studies on the pairing problem, on algebraic models and on the Brueckner theory are also actively pursued.

\section*{Acknowledgments}
We wish to thank Prof. G. Marchesini, head of the national INFN Theory Group, for supporting the institution of the PI32 collaboration.

\end{document}